\documentclass[conference]{IEEEtran}
\IEEEoverridecommandlockouts
\usepackage{cite}
\usepackage{amsmath,amssymb,amsfonts}
\usepackage{algorithmic}
\usepackage{graphicx}
\usepackage{textcomp}
\usepackage{makecell}
\usepackage{xcolor}
\usepackage{svg}
\usepackage{listings}
\usepackage{url}
\usepackage{framed}
\usepackage{caption}
\usepackage{tabularx}

\def\BibTeX{{\rm B\kern-.05em{\sc i\kern-.025em b}\kern-.08em
    T\kern-.1667em\lower.7ex\hbox{E}\kern-.125emX}}

\makeatletter
\def\endthebibliography{%
  \def\@noitemerr{\@latex@warning{Empty `thebibliography' environment}}%
  \endlist
}
\makeatother

\usepackage{color}
\usepackage{courier}
\usepackage{sourcecodepro}
\usepackage[T1]{fontenc}

\lstdefinestyle{promptstyle}{
  showspaces=false,                
  basicstyle=\ttfamily\footnotesize,
  breaklines=true,                   
  tabsize=1,                         
  breakindent=0pt,
}

\lstdefinestyle{larkstyle}{
  language=Python, 
  basicstyle=\ttfamily\footnotesize,
  breaklines=true,
  columns=fullflexible,
  morekeywords={GLOBAL, LLEFT, NUM, RRIGHT, ASSIGN, LPAREN, RPAREN, SEMI, ID},  
  showstringspaces=false,
}

\begin{document}

\title{Prose-to-P4: Leveraging High Level Languages}

\author{
\IEEEauthorblockN{Mihai-Valentin Dumitru, Vlad-Andrei Bădoiu, Costin Raiciu}
\IEEEauthorblockA{University Politehnica of Bucharest\\
\{mihai.dumitru2201, vlad\_andrei.badoiu, costin.raiciu\}@upb.ro}}

\maketitle

\begin{abstract}
Languages such as P4 and NPL have enabled a wide and diverse range of
networking applications that take advantage of programmable dataplanes.
However, software development in these languages is difficult.
To address this issue, high-level languages have been designed to offer
programmers powerful abstractions that reduce the time, effort and
domain-knowledge required for developing networking applications.
These languages are then translated by a compiler into P4/NPL code.

Inspired by the recent success of Large Language Models (LLMs) in the task of
code generation, we propose to raise the level of abstraction even higher,
employing LLMs to translate prose into high-level networking code.
We analyze the problem, focusing on the motivation and opportunities, as well as
the challenges involved and sketch out a roadmap for the development of a system
that can generate high-level dataplane code from natural language instructions.
We present some promising preliminary results on generating Lucid code from
natural language.
\end{abstract}


\section{Introduction}%
\label{sect:intro}

\begin{figure*}[t]
\centering
\includesvg[width=0.9\textwidth]{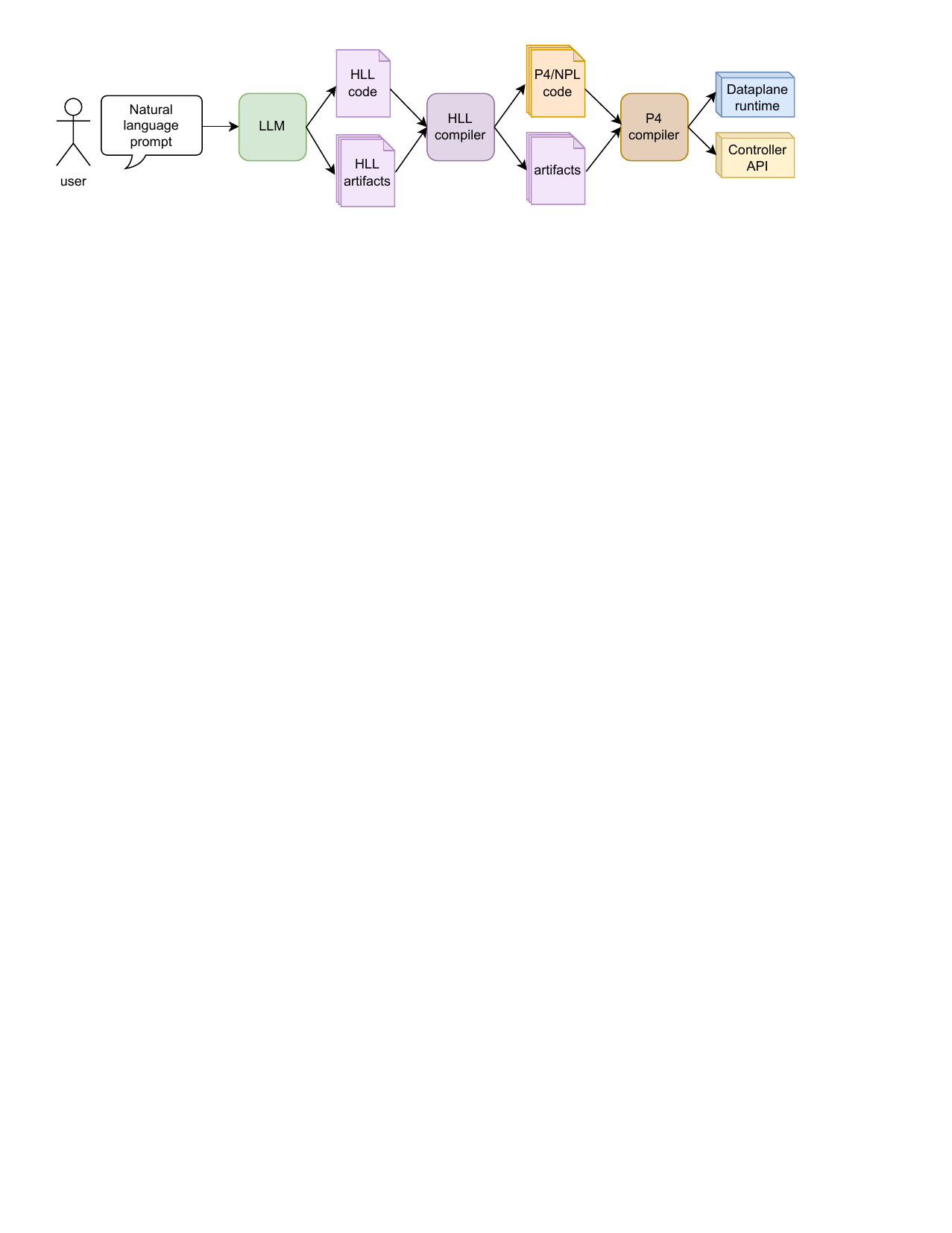}
\caption{Proposed pipeline of going from prose to concrete dataplane
applications on actual switches.
Some languages may require network configurations or other extra information,
grouped here under ``input artifacts''.
Compilation produces one or more dataplane programs, perhaps together with
program-to-device mappings or other information, grouped here as ``output
artifacts''.}
\label{fig:flow}
\end{figure*}

The introduction of programmable dataplanes and associated languages, such as
P4~\cite{bosshart2014p4} and NPL~\cite{npl}, has opened the door for a broad
range of networking applications~\cite{hauser2023survey}.

Software development in these languages has proven to be
difficult~\cite{gao2020lyra,loehr2024lucid}.
For example, limited hardware resources require programmers to be familiar with
the target and customize programs for it, reducing portability; adding or
removing support for a protocol requires changes in multiple parts of the
program (parser, deparser, control), making it hard to compose programs or to
compartmentalize functionality into separate libraries.

One strategy to address these problems is the development of high-level
dataplane programming languages (HLDPLs) that compile to P4/NPL code for one or
more switches, enabling programmers themselves to not focus on low-level
details.
Over the years, quite a few such languages have been designed, including
Graph-to-P4~\cite{zaballa2019graph},
Lucid~\cite{sonchack2021lucid,loehr2024lucid},
Lucid 2.0~\cite{loehr2022safe},
Lyra~\cite{gao2020lyra},
O4~\cite{alcoz2022reducing},
P4All~\cite{hogan2020elastic},
P4rrot~\cite{gyorgyi2023p4rrot},
pcube~\cite{shah2018pcube}.
These are much closer to natural language, provide less burden on the developer
and in many cases have a less complex grammar.

The field remains volatile: new languages appear periodically and none of the
existing ones has been widely adopted and given the opportunity to mature.

Inspired by the recent success of Large Language Models (LLMs) in the task of
code
generation~\cite{chen2021evaluating,wei2023magicoder,li2023starcoder,lozhkov2024starcoder,guo2024deepseek,team2023gemini},
we wish to push the move towards higher-level network programming languages even
further, by employing natural language as the top layer for the development of
dataplanes.
Ideally, the network programmer should be able to express desired
functionalities in the form of a natural language prompt that is then
progressively transformed into code that runs on actual devices (Figure
\ref{fig:flow}).
The user of such a system can not only ignore platform-specific details, but can
do away with learning to program in a language that may easily be abandoned or
superseded.
All that is needed is a rudimentary knowledge of basic networking architectures
and protocols.

One of the main obstacles in developing such a system is the low amount of data.
The authors of The Stack v2~\cite{lozhkov2024starcoder}, one of the largest
publicly available coding datasets, refer to Perl as a ``low-resource
language'' -- with over a million documents available, totalling 7.82 GB.
We wish to address ``no-resource'' languages, for which only several
programs are available.

Our focus is not code-generation for a particular, already existing dataplane
language, but rather code-generation for recently published, non-mature HLDPLs,
for which there is no existing program dataset or active community.

One may ask what makes dataplane networking a good candidate for code generation
under the no-resources constraints. Computations intended to be run in the
dataplane are linear in nature; each program can be modeled as a chain of
simple instructions, with occasional branching for conditional blocks, but no
loops.
Due to the nature of the applications themselves, as well as the constraints
ultimately mandated by the target switches, programs are also quite limited in
length.
We believe these features make HLDPLs ideal candidates for using LLMs to
generate ``no-resource languages''.

The aim of this work is to leverage code-generating LLMs and conventional
transpilers to transform natural languages prompts into concrete code that can
run on a programmable switch.
The opportunity to do so is enabled by the simple nature of HLDPLs and the
relatively short program length.
The main challenge stems from the lack of a dataset of programs written in these
languages: code-generating LLMs are usually trained or fine-tuned on gigabytes,
or even terabytes of data; we only assume the existence of a handful of
examples.

We make the following key contributions:

\begin{itemize}
  \item identify a series of characteristics that make HLDPLs
    good candidates for no-resource code generation
  \item analyze the challenges and opportunities of utilizing LLMs to generate
      no-resource languages
  \item evaluate the plausibility of our methodology using the Lucid
      language~\cite{sonchack2021lucid} as a target
\end{itemize}

\section{High-level dataplane programming languages}%
\label{sect:languages}

The difficult nature of software development in P4 has lead to research into
new dataplane programming languages which abstract away target-specific details,
offering programmers high-level constructs to aid with portability,
composability etc.
These language differ in the level of abstraction, scope and motivation but in
general they aim to reduce time, effort and domain-knowledge needed to develop
new dataplane applications.

We briefly survey existing languages, focusing on features relevant to our goal
of LLM-assisted code generation and present our rationale for singling out
Lucid~\cite{sonchack2021lucid} for our preliminary experiments.

\textbf{O4}~\cite{alcoz2022reducing}, \textbf{pcube}~\cite{shah2018pcube},
\textbf{P4All}~\cite{hogan2020elastic} add to P4 several syntactical constructs
such as fixed-size loops and arrays (O4, pcube) and \emph{elastic structures}
(P4All).
This results in languages syntactically similar to P4 itself.

\textbf{Graph-to-P4}~\cite{zaballa2019graph} can translate state diagrams
(designed visually and represented as an XML) into parser graphs.
Its scope is limited to P4 parsers and does not address other components of
dataplane programming, such as control blocks.

\textbf{P4rrot}~\cite{gyorgyi2023p4rrot} aims to ease the implementation of
``application-layer tasks'' in the dataplane.
It is designed as a Python library, which has interesting implications,
as Python seems to be at the center of LLM-based code generation efforts.
For the purposes of this work, we choose to focus on a language with a
standalone syntax, as this is more common among the languages considered.

\textbf{Lyra}~\cite{gao2020lyra} addresses the issues of portability,
extensibility and composition.
The Lyra compiler takes as input code in the Lyra language, an
algorithm scope describing the placement of various and a description of the
network's topology and configuration; from these it first produces a
``context-aware intermediate representation'' that can be then turned by the
backend into P4 or NPL code.
Unfortunately neither a complete grammar, a compiler implementation, detailed
documentation or complete examples are publicly available.

\textbf{Lucid}~\cite{sonchack2021lucid}'s goal is ``putting control
functionality into the data plane.''
Lucid introduces events (such as an arriving packet) and event handlers --
procedures that describe stateful computation to be executed when an event
occurs, as well as a novel type system and memory operations.
We consider Lucid the most mature HLDPL currently available.
It has been followed by \textbf{Lucid
2.0}~\cite{loehr2022safe}, which introduces new constructs to aid with modular
  programming, such as polymorphism and type inference.
The presentation has also been extended to a PhD thesis~\cite{loehr2024lucid}
that provides additional detail on the language design and motivation.
Both the compiler and example programs are openly
available~\footnote{\url{https://github.com/PrincetonUniversity/lucid}}; this
makes Lucid an ideal candidate for our experiments, because it offers all the
resources that we may expect.

At the time of writing, none of these languages has been standardised or
widely-adopted; they are research projects in various degrees of active
development.
We expect this volatility to continue in the near future, with existing
languages being improved or modified and new languages being designed.
The lack of stability in the field makes it difficult for novices (and perhaps
experienced programmers alike) to enjoy the benefits of these languages, since
the learning effort can be considered wasted if a language is soon
abandoned by the community or superseded by a new one.

We wish to offer potential network programmers the possibility of rapid
prototyping with low effort and little domain-knowledge, harnessing the power of
HLDPLs without requiring the commitment to familiarize oneself
with a particular language: its syntax, semantics and idiosyncrasies.
To this end, we propose using LLMs to translate natural language specifications
into code.

\section{Generating no-resource languages}%
\label{sect:generating}

Our goal is to use LLMs to translate natural language specifications into HLDPL
code.
Due to their maturity and impressive results on a wide range of applications, we
target ChatGPT 4 and Gemini Ultra. 
While LLMs have, in recent times, shown very good results in the task of
prose-to-code, they usually generate code in popular programming languages for
which a large number of programs written by human developers is available;
models are fine-tuned or trained on these large code datasets.
For a newly published programming language with no active community, such a
wide range of examples is not available, making fine-tuning or training
non-viable strategies.

We assume the resources available are a subset of the following:

\begin{itemize}

    \item the contents of the scientific paper that introduces the language

    \item additional documentation

    \item description of the formal grammar

    \item several code examples

    \item a compiler implementation

\end{itemize}

The documentation, grammar and examples could be available as appendices to the
original paper, or as part of a public repository; a compiler could be useful
for the possibility of extracting a formal grammar description from its parser,
as well as for offering a better understanding of the semantics of the language.

Lacking a dataset of programs to retrain the LLM, we have to rely on
\emph{prompt engineering} techniques, which involve carefully constructing the
inputs to the LLM, guiding it to produce high-quality answers.
Prompt engineering has already developed into a wide and diverse
field~\cite{sahoo2024systematic} with techniques such as
Chain-of-Thought~\cite{wei2022chain} and Tree-of-thoughts~\cite{yao2024tree}
achieving impressive results on tasks such as arithmetic reasoning, symbolic
reasoning, creative writing etc.

The technique of ``grammar prompting'' developed by Wang et
al.~\cite{wang2024grammar} is most relevant to our goals.
The authors tackle the problem of generating programs in a Domain Specific
Language (DSL) that is absent (or present in small quantities) in the LLM's
training set.
To this end, a few-shot approach is employed, attaching to each example a small
subset of the DSL grammar, ``minimally sufficient'' for generating that
particular example.
The model is then required to first produce a specialized grammar for the
requested program, then generate the code conditioned on this grammar.

That same paper presents state-of-the-art results on five DSLs for the tasks of
semantic parsing, AI planning and molecule generation.
Jain et al.~\cite{jain2023generating} experiment with grammar prompting, as well
as other techniques, for the task of prose-to-diagram: translating a natural
language description into the syntax used by the Penrose framework, which
produces visual diagrams.

Because our goal is to develop a system for generating newly published
languages, it is relevant to know whether data associated with the tested
language (Lucid), such as the text of the original paper or the public
repository code, are part of these LLMs' training sets.
Unfortunately, for ChatGPT 4 and Gemini Ultra, this information is not publicly
available; basic interrogation of the two LLMs reveal that they have at
least some knowledge of Lucid.
This is a limitation of our experiments and results, because the language
information does not come solely from the contents of the prompt, which may cast
doubt on whether the results could transfer to completely new programming
languages, completely absent from the training dataset.

However, we illustrate that this is not a critical limitation, by building a
baseline of results that show how, without advanced prompt-engineering
techniques, neither ChatGPT 4 nor Gemini Ultra can generate code in the
considered languages.

\section{Preliminary results}%
\label{sect:results}

We evaluate the grammar prompting technique~\cite{wang2024grammar} for few-shot
HLDPL code generation in Lucid, because of its standalone syntax and public
availability of all the resources mentioned in \S \ref{sect:generating}.
To this end, we leverage the Gemini and ChatGPT models due to their overall
performance on a wide variety of tasks, including natural language
understanding, symbolic reasoning and code generation, which we believe are
useful skills for the task at hand.

%
%
%


One important constraint for prompt engineering is the \emph{context length} of
the LLM -- the number of tokens that the LLM can effectively consider when
producing the answer.
A naive prompt for Lucid, consisting of task-specific instructions, the contents
of the original paper~\cite{sonchack2021lucid}, a complete grammar in
Backus-Naur form (extracted from the publicly available parser) and the full
code of the ten applications presented in the paper has a length of 35K tokens;
this is more than the web interfaces of both ChatGPT 4 and Gemini Ultra can
handle.
However, state of the art models have context lengths in the hundreds of
thousands, even millions of tokens; Google's Gemini 1.5~\cite{reid2024gemini}
boasts of a context length of ten million tokens.
We ran only a few simple tests using OpenAI's API to query GPT 4.
The results seem syntactically coherent but more testing is needed.
For the rest of our experiments, we used the web interface of ChatGPT and
Gemini (to avoid API costs) and employed grammar prompting.

To integrate Lucid into the framework, we wrote a grammar for it in
Lark~\footnote{\url{https://github.com/lark-parser/lark}} and
designed 20 samples for few-shot learning. The samples vary from basic concepts (e.g.
defining the type for IPv4 packets), to more complex prompts (i.e. MAC address
learning).
Lastly, we modify the base prompt to introduce a high level description of the
HLDPL.

Besides the program written in the Lucid language, a P4 template is also
required, containing basic elements such as parsers and deparser, as well as
annotations for where the high-level bits should fit.
For the purposes of this paper, we do not require the LLM to generate it.

The resulting grammar-learning prompts are structured as shown in Figure~\ref{fig:prompt}.
They sum up to around 5.5K tokens; the responses are just several hundred tokens
in length.

The model is provided with the full Lucid grammar and several
example queries consisting of a specialized grammar (subset of the complete
grammar) and the resulting code that
implements the query.
Lastly, the model is asked to predict a grammar for a given query and generate
the program that implements a query conforming to it
(Figure~\ref{fig:response}).

We have tested simple queries, requiring one-line responses and three complex
ones: writing a firewall that can drop packets based on the
connection 5-tuple (partial answer in Figure~\ref{fig:response}), a stateful NAT
and a 6to4 tunnel.

From our experiments, we noticed that Gemini sometimes fails to conform to the
requirements and does not produce a grammar subset to drive the generated code,
resulting in incorrect syntax even for simple examples.

ChatGPT 4 is able to follow the guidance of grammar prompting and always produces
a grammar for its output; however, this grammar is not always correct, so
neither is the accompanying program.
Figure~\ref{fig:response} shows a fragment of ChatGPT 4's answer for a firewall
query.

\begin{figure}
\begin{framed}
  \textbf{Prompt to LLM}
  \begin{lstlisting}[language={},style=promptstyle]
You are an expert dataplane programmer, and you need to write a program for the given language query.
A complete grammar of the language is presented below. First, you should write grammar rules by choosing from the following BNF rules.
Then, you should write programs that conform to your predicted rules.

[BEGIN RULES]
// Full definition of the Lucid grammar
[END RULES]

query: generate a global array "nat_to_port" with NUM_PORTS values
\end{lstlisting}

\begin{lstlisting}[style=larkstyle]
BNF grammar rules:
prog ::= decls
decls ::= decl
decl ::= GLOBAL cid poly ID ASSIGN cid paren_args SEMI
GLOBAL ::= "global"
cid ::= ID "." cid | ID
ID ::= "Array" | "t" | "nat_to_port" | "create" | "NUM_PORTS"
poly ::= LLEFT polys RRIGHT
LLEFT ::= "<<"
polys ::= size
size ::= NUM
NUM ::= "32"
RRIGHT ::= ">>"
ASSIGN ::= "="
paren_args ::= LPAREN args RPAREN
LPAREN ::= "("
args ::= exp
exp ::= cid
RPAREN ::= ")"
SEMI ::= ";"
\end{lstlisting}

\begin{lstlisting}[style=promptstyle]
program based on the BNF grammar rules:
global Array.t<<32>> nat_to_port  = Array.create(NUM_PORTS);
\end{lstlisting}

\begin{lstlisting}[language={},style=promptstyle]
// other examples

query: <the desired query>
\end{lstlisting}

\end{framed}
\captionof{figure}{Sample of prompt engineering for Lucid}
\label{fig:prompt}
\end{figure}

\begin{figure}
\begin{framed}

\begin{lstlisting}[style=promptstyle]
BNF grammar rules:
\end{lstlisting}

\begin{lstlisting}[style=larkstyle]
prog ::= decls
decls ::= decl decls | decl
decl ::= "const" ty ID ASSIGN exp SEMI | GLOBAL cid poly ID ASSIGN cid paren_args SEMI | event_decl SEMI | HANDLE ID paramsdef "{" statement "}" | FUN ty ID paramsdef "{" statement "}"
ty ::= TINT | TINT single_poly | TBOOL
[...]
FUN ::= "fun"
RETURN ::= "return"

program based on the BNF grammar rules:

const int HASH_SEED = 1234;
const int NUM_PORTS = 6;
global Array.t<<32>> nat_to_ip = Array.create(NUM_PORTS);
global Array.t<<32>> nat_to_port = Array.create(NUM_PORTS);

[...]

handle outside_packet(int<<32>> dst_port) {
    int ip = Array.get(nat_to_ip, dst_port);
    int<<32>> port = Array.get(nat_to_port, dst_port);
    generate outside_continue(ip, port);
}
\end{lstlisting}

\end{framed}
\captionof{figure}{ChatGPT response for the query: ``a stateful NAT''.}
\label{fig:response}
\end{figure}

Table~\ref{tab:results} summarizes ChatGPT's performance on the three complex
tasks; for each task, it lists the number of rules in the output grammar
(``\texttt{A ::= B | C}'' counts as two rules), the number of lines of code in
the output Lucid program and the number of lines that need to be fixed for the
program to compile.
The task prompts are admittedly vague, so it is hard to assess the
``correctness'' of the implementation, which is why we focus on
whether programs compile.
The generated code is aligned with the input query and could serve well as a
starting point for a more particular implementation.

While always generating a grammar fragment, ChatGPT doesn't always respect it.
For the tunnel implementation, it uses hexadecimal constants with the ``0x''
prefix, which are invalid both in the fragment generated and in the Lucid
grammar; it also uses a bitwise ``\&'' without generating rules for it.
Conversely, it includes rules that it does not use; for example it allows the
equality operator ``=='' then generates a program that does not perform any equality check.

\begin{table}
\footnotesize
\centering
\begin{tabularx}{\linewidth}{X X X X}
 \hline
 \textbf{Task} & \textbf{\# of rules} & \textbf{LoC} & \textbf{LoC to fix} \\
 \hline\hline
 Firewall & 91 & 17 & 2  \\
 NAT & 97 & 39 & 0 \\
 6to4 tunnel & 74 & 15 & 4   \\
 \hline
\end{tabularx}
\caption{Summary of the results on the three complex tasks}
\label{tab:results}
\vspace{-0.2cm}
\end{table}

\section{Discussion and Future Work}%
\label{sect:future}

The results presented in \S \ref{sect:results} are just a initial move towards
the task of translating natural language instructions into HLDPLs.
An immediate next step is mostly quantitative in nature: increasing the number
of example programs included in the prompt and running additional tests to
generate snippets and programs of increasing complexity.
We would like to do further experiments both with the baseline prompt (a large
dump of all available resources) and with grammar prompting, so that we can have
evaluation metrics that allow us to numerically compare the efficacy of the two
methods.

The issues mentioned at the end of \S \ref{sect:results}, concerning constructs
that are not valid for the grammar fragment and grammar rules that are unneeded,
warrant additional attention.
We believe they can be addressed effectively through a loop of additional
queries that point out the errors in the output.

Lucid requires a P4 harness with the basic structure of the dataplane,
together with annotations for where the functionality described in the
high-level language should fit.
In our tests, we only focused on the Lucid language itself; in the future, we
would like the LLM to also generate the P4 template.

We also wish to explore code generation in P4rrot, whose very different
structure can cast light on whether our technique generalizes to other
languages.

Information about Lucid, as well as about all the languages presented in \S
\ref{sect:languages}, is very likely part of the dataset on which ChatGPT 4 and
Gemini were trained.
Our experiments show that grammar prompting yields better results than the naive
baseline of relying solely on the LLM's existing knowledge, but
such improvements may be possible solely due to the latent knowledge of the LLM.
To explore this issue, we must experiment with new languages that are not part
of the training dataset.
We plan to design a minimalistic HLDPL, producing all the relevant artifacts
(description of the language, formal grammar, code examples) and apply the
techniques presented here to evaluate code generation for this new language.

\section{Conclusion}%
\label{sect:conclusions}

In order to address the difficulties of programming in P4 or NPL, several
high-level dataplane programming languages have been proposed.
As this is an active, ever-changing field, neither of these have been
standardized or widely-adopted.

Thanks to the simple structure, narrow scope and short program length,
we believe these languages are good candidates for the task of employing LLMs to
generate languages for which virtually no dataset is available.

In \S \ref{sect:results} we presented preliminary results of using ChatGPT
4 and Gemini in order to generate code for the Lucid language.

\bibliographystyle{ieeetr}
\bibliography{biblio}

\begin{thebibliography}{10}

\bibitem{bosshart2014p4}
P.~Bosshart, D.~Daly, G.~Gibb, M.~Izzard, N.~McKeown, J.~Rexford,
  C.~Schlesinger, D.~Talayco, A.~Vahdat, G.~Varghese, {\em et~al.}, ``P4:
  Programming protocol-independent packet processors,'' {\em ACM SIGCOMM
  Computer Communication Review}, vol.~44, no.~3, pp.~87--95, 2014.

\bibitem{npl}
Broadcom, ``{NPL: Open, High-Level language for developing feature-rich
  solutions for programmable networking platforms},'' 2019.

\bibitem{hauser2023survey}
F.~Hauser, M.~H{\"a}berle, D.~Merling, S.~Lindner, V.~Gurevich, F.~Zeiger,
  R.~Frank, and M.~Menth, ``A survey on data plane programming with p4:
  Fundamentals, advances, and applied research,'' {\em Journal of Network and
  Computer Applications}, vol.~212, p.~103561, 2023.

\bibitem{gao2020lyra}
J.~Gao, E.~Zhai, H.~H. Liu, R.~Miao, Y.~Zhou, B.~Tian, C.~Sun, D.~Cai,
  M.~Zhang, and M.~Yu, ``Lyra: A cross-platform language and compiler for data
  plane programming on heterogeneous asics,'' in {\em Proceedings of the Annual
  conference of the ACM Special Interest Group on Data Communication on the
  applications, technologies, architectures, and protocols for computer
  communication}, pp.~435--450, 2020.

\bibitem{loehr2024lucid}
D.~K. Loehr {\em et~al.}, ``Lucid: A high-level, easy-to-use dataplane
  programming language,'' 2024.

\bibitem{zaballa2019graph}
E.~O. Zaballa and Z.~Zhou, ``Graph-to-p4: A p4 boilerplate code generator for
  parse graphs,'' in {\em 2019 ACM/IEEE Symposium on Architectures for
  Networking and Communications Systems (ANCS)}, pp.~1--2, IEEE, 2019.

\bibitem{sonchack2021lucid}
J.~Sonchack, D.~Loehr, J.~Rexford, and D.~Walker, ``Lucid: A language for
  control in the data plane,'' in {\em Proceedings of the 2021 ACM SIGCOMM 2021
  Conference}, pp.~731--747, 2021.

\bibitem{loehr2022safe}
D.~Loehr and D.~Walker, ``Safe, modular packet pipeline programming,'' {\em
  Proceedings of the ACM on Programming Languages}, vol.~6, no.~POPL,
  pp.~1--28, 2022.

\bibitem{alcoz2022reducing}
A.~G. Alcoz, C.~Busse-Grawitz, E.~Marty, and L.~Vanbever, ``Reducing p4
  language's voluminosity using higher-level constructs,'' in {\em Proceedings
  of the 5th International Workshop on P4 in Europe}, pp.~19--25, 2022.

\bibitem{hogan2020elastic}
M.~Hogan, S.~Landau-Feibish, M.~Tahmasbi~Arashloo, J.~Rexford, D.~Walker, and
  R.~Harrison, ``Elastic switch programming with p4all,'' in {\em Proceedings
  of the 19th ACM Workshop on Hot Topics in Networks}, pp.~168--174, 2020.

\bibitem{gyorgyi2023p4rrot}
C.~Gy{\"o}rgyi, S.~Laki, and S.~Schmid, ``P4rrot: Generating p4 code for the
  application layer,'' {\em ACM SIGCOMM Computer Communication Review},
  vol.~53, no.~1, pp.~30--37, 2023.

\bibitem{shah2018pcube}
R.~Shah, A.~Shirke, A.~Trehan, M.~Vutukuru, and P.~Kulkarni, ``pcube:
  Primitives for network data plane programming,'' in {\em 2018 IEEE 26th
  International Conference on Network Protocols (ICNP)}, pp.~430--435, IEEE,
  2018.

\bibitem{chen2021evaluating}
M.~Chen, J.~Tworek, H.~Jun, Q.~Yuan, H.~P. d.~O. Pinto, J.~Kaplan, H.~Edwards,
  Y.~Burda, N.~Joseph, G.~Brockman, {\em et~al.}, ``Evaluating large language
  models trained on code,'' {\em arXiv preprint arXiv:2107.03374}, 2021.

\bibitem{wei2023magicoder}
Y.~Wei, Z.~Wang, J.~Liu, Y.~Ding, and L.~Zhang, ``Magicoder: Source code is all
  you need,'' {\em arXiv preprint arXiv:2312.02120}, 2023.

\bibitem{li2023starcoder}
R.~Li, L.~B. Allal, Y.~Zi, N.~Muennighoff, D.~Kocetkov, C.~Mou, M.~Marone,
  C.~Akiki, J.~Li, J.~Chim, {\em et~al.}, ``Starcoder: may the source be with
  you!,'' {\em arXiv preprint arXiv:2305.06161}, 2023.

\bibitem{lozhkov2024starcoder}
A.~Lozhkov, R.~Li, L.~B. Allal, F.~Cassano, J.~Lamy-Poirier, N.~Tazi, A.~Tang,
  D.~Pykhtar, J.~Liu, Y.~Wei, {\em et~al.}, ``Starcoder 2 and the stack v2: The
  next generation,'' {\em arXiv preprint arXiv:2402.19173}, 2024.

\bibitem{guo2024deepseek}
D.~Guo, Q.~Zhu, D.~Yang, Z.~Xie, K.~Dong, W.~Zhang, G.~Chen, X.~Bi, Y.~Wu,
  Y.~Li, {\em et~al.}, ``Deepseek-coder: When the large language model meets
  programming--the rise of code intelligence,'' {\em arXiv preprint
  arXiv:2401.14196}, 2024.

\bibitem{team2023gemini}
G.~Team, R.~Anil, S.~Borgeaud, Y.~Wu, J.-B. Alayrac, J.~Yu, R.~Soricut,
  J.~Schalkwyk, A.~M. Dai, A.~Hauth, {\em et~al.}, ``Gemini: a family of highly
  capable multimodal models,'' {\em arXiv preprint arXiv:2312.11805}, 2023.

\bibitem{sahoo2024systematic}
P.~Sahoo, A.~K. Singh, S.~Saha, V.~Jain, S.~Mondal, and A.~Chadha, ``A
  systematic survey of prompt engineering in large language models: Techniques
  and applications,'' {\em arXiv preprint arXiv:2402.07927}, 2024.

\bibitem{wei2022chain}
J.~Wei, X.~Wang, D.~Schuurmans, M.~Bosma, F.~Xia, E.~Chi, Q.~V. Le, D.~Zhou,
  {\em et~al.}, ``Chain-of-thought prompting elicits reasoning in large
  language models,'' {\em Advances in neural information processing systems},
  vol.~35, pp.~24824--24837, 2022.

\bibitem{yao2024tree}
S.~Yao, D.~Yu, J.~Zhao, I.~Shafran, T.~Griffiths, Y.~Cao, and K.~Narasimhan,
  ``Tree of thoughts: Deliberate problem solving with large language models,''
  {\em Advances in Neural Information Processing Systems}, vol.~36, 2024.

\bibitem{wang2024grammar}
B.~Wang, Z.~Wang, X.~Wang, Y.~Cao, R.~A~Saurous, and Y.~Kim, ``Grammar
  prompting for domain-specific language generation with large language
  models,'' {\em Advances in Neural Information Processing Systems}, vol.~36,
  2024.

\bibitem{jain2023generating}
R.~Jain, W.~Ni, and J.~Sunshine, ``Generating domain-specific programs for
  diagram authoring with large language models,'' in {\em Companion Proceedings
  of the 2023 ACM SIGPLAN International Conference on Systems, Programming,
  Languages, and Applications: Software for Humanity}, pp.~70--71, 2023.

\bibitem{reid2024gemini}
M.~Reid, N.~Savinov, D.~Teplyashin, D.~Lepikhin, T.~Lillicrap, J.-b. Alayrac,
  R.~Soricut, A.~Lazaridou, O.~Firat, J.~Schrittwieser, {\em et~al.}, ``Gemini
  1.5: Unlocking multimodal understanding across millions of tokens of
  context,'' {\em arXiv preprint arXiv:2403.05530}, 2024.

\end{thebibliography}

\end{document}